# Topological surface states in the Kondo insulator YbB$_{12}$ revealed via planar tunneling spectroscopy

A. Gupta[1,2], A. Weiser[3], L. H. Greene[1,2], L. Pressley[4,†], Y. Luo[4], C. Lygouras[4], J. Trowbridge[5], W.A. Phelan[4,5,6,‡], C.L. Broholm[4], T. McQueen[4,5,7], W. K. Park[1,*]

[1]National High Magnetic Field Laboratory, Florida State University, Tallahassee, FL 32310, USA

[2]Department of Physics, Florida State University, Tallahassee, FL 32306, USA

[3]Department of Physics, Astronomy, Geology, and Environmental Science, Youngstown State University, Youngstown, OH 44555, USA

[4]Department of Physics and Astronomy, Johns Hopkins University, Baltimore, MD 21218, USA

[5]Department of Chemistry, Johns Hopkins University, Baltimore, MD 21218, USA

[6]Hopkins Extreme Materials Institute, Johns Hopkins University, Baltimore, MD 21218, USA

[7]Department of Materials Science and Engineering, Johns Hopkins University, Baltimore, MD 21218, USA

[*]Corresponding author: wkpark@magnet.fsu.edu

[†]Present address: Oak Ridge National Laboratory, Oak Ridge, TN 37830, USA

[‡]Present address: Los Alamos National Laboratory, Los Alamos, Mail Stop E574, Los Alamos, NM 87545, USA

**Abstract**

Planar tunneling spectroscopy of the Kondo insulator $SmB_6$ suggests that an interaction between the surface Dirac fermions and the bulk spin excitons results in incompletely protected topological surface states. To gain further insight into their true nature, it is necessary to study other topological Kondo insulator candidates. Calculations of electronic energy bands predict that the Kondo insulator $YbB_{12}$ hosts topological surface states protected by crystalline mirror symmetry. In this study, we present tunneling conductance spectra obtained from the (001) surface of $YbB_{12}$ single crystals and discuss them in comparison to $SmB_6$. The linear conductance at low bias provides strong evidence for the existence of surface Dirac fermions. The double-hump structure in the negative bias region is associated with hybridized band edges, in agreement with a calculated band structure. While these similarities with $SmB_6$ are suggestive of the existence of topological surface states in $YbB_{12}$, in agreement with other experiments, some discrepancies are also observed, which we attribute to a difference in their exact nature from those in $SmB_6$.

I.  INTRODUCTION

The remarkable discovery that band insulators can be classified into topologically distinct classes by virtue of their topological indices gave birth to the exciting field of topological insulators (TIs) [1,2]. The transition from a topologically trivial to a nontrivial phase of the same insulator is not associated with the breaking of any symmetry and thus cannot be understood within the Ginzburg-Landau picture [3]. Such a topological phase transition can only occur by closing and reopening the band gap, which gives rise to conducting surface states (SS) protected by spin-momentum locking [4]. The nature of these SS heavily depends on the strength of the electron correlations in each material. Therefore, the SS in weakly correlated *s* and *p* orbital systems like HgTe [5-7] and the $Bi_2Se_3$ family of compounds [8-10] are less complex to understand than those in *d* and *f* orbital systems because of the added ingredient of strong correlations [11]. Kondo insulators (KIs) are one such class of strongly correlated materials, in which the hybridization between the conduction band and the localized states results in the opening of a hybridization gap around the Fermi level at low temperature [12,13]. The hybridized band edges are brought closer due to the strong correlation effect, resulting in a reduced gap [12,14]. In combination with the large spin-orbit interaction, this can cause band inversions in these compounds. Due to the odd parity of *f*-states, band inversions at odd number of high-symmetry points can give rise to a topologically nontrivial phase, known as topological Kondo insulators (TKIs) [14].

In the KI $SmB_6$, such a hybridization takes place between the 4*f* orbitals and 5*d* bands, resulting in a nontrivial $Z_2$ topological phase protected by time-reversal symmetry [1,2] and hence making it a prime TKI candidate [11,15]. This theoretical prediction has stimulated extensive studies of $SmB_6$. The low-temperature resistance plateau, which was repeatedly observed in various transport measurements, has been attributed to the topologically protected SS [16-18]. While the observation of de Haas-van Alphen (dHvA) effect in $SmB_6$ suggests the presence of Fermi surfaces in $SmB_6$ [19-22], the absence of Shubnikov-de Haas (SdH) oscillations remains a mystery. Furthermore, the exact origin of the dHvA oscillations in $SmB_6$ is still under debate [19-22]. The hybridization gap has been detected in angle-resolved photoemission [23-29], tunneling [30-34], and point-contact spectroscopic [35] studies. Our recent planar tunneling spectroscopy (PTS) studies on $SmB_6$ have provided clear signatures for surface Dirac fermions [33], in good agreement with electronic energy band calculations [14,36-38] and a quantum oscillation study [19]. Moreover, the topological surface states (TSS) were found to interact with bulk spin excitons (SEs) [39], in agreement with a theoretical prediction [40] and an ARPES study [41]. Such an interaction would result in incomplete protection of the TSS. Even though the helical spin texture of the SS of $SmB_6$ was reported in some measurements [29,42,43], it still requires further investigations particularly considering that the topological nature of the $SmB_6$ SS has been debated [15,44]. This motivates us to study another TKI candidate to better understand the origin of the metallic SS in these materials.

Ytterbium dodecaboride ($YbB_{12}$) is another prototypical KI, which has a face-centered cubic structure with the lattice constant of 7.468 Å (Fig. 1a), much larger than that of $SmB_6$ (4.134 Å) [45]. The calculated band structure of $YbB_{12}$ differs from $SmB_6$ in that the band inversion occurs twice at each of the three $X$ points (Fig.1b) [46], resulting in a trivial $Z_2$ invariant. Instead, band calculations for the (001) surface [46] show that $YbB_{12}$ is a topological crystalline insulator [47,48] with a nontrivial mirror Chern number protected by crystalline reflection symmetry with respect to the $\Gamma X_1 X_2$ mirror plane. Transport measurements on $YbB_{12}$ single crystals reveal two activation-type energy gaps, 11.7 meV (below 40 K) and 3.8 meV (below 15 K) [49], but their origin is yet to be understood. $YbB_{12}$ shows a low-temperature resistivity plateau similar to $SmB_6$, which can be attributed to the TSS. The weak antilocalization effect displayed in the magnetoresistance of $YbB_{12}$ microstructures has been argued to arise from the spin-momentum locking property of the surface states [49]. Quantum oscillations have been observed in $YbB_{12}$ arising from both dHvA and SdH effects [49-52]. The angular variation of the SdH period was found to point to a three-dimensional Fermi surface [50] but such oscillations were not observed in the micro-structured $YbB_{12}$ single crystals [49], both of which indicate that these oscillations originate from the electrically insulating bulk. This is rather puzzling since the insulating bulk is not expected to contain a Fermi surface. On the other hand, the origin of the dHvA oscillations in $YbB_{12}$ is under debate [50,52]. The formation of the hybridization gap was observed in optical conductivity [53] and photoemission spectroscopy measurements [54,55]. The optical conductivity measurements found the size of the indirect (direct) gap to be 15 meV (~180 meV) [53]. Studies based on inelastic neutron scattering have found in-gap spin excitations at two energy scales, 14.5 meV and 20 meV [56], but a Lu-substitution study has suggested that a more local phenomenon is responsible for the formation of the gap [57]. The hybridization gap can be closed at high magnetic fields in the range of 45-47 T ($H \parallel$ [100]) and 55-59 T ($H \parallel$ [110]), turning the system into a metallic state [51,52,58-60].

However, transport measurements alone are not enough to elucidate the topological nature of the SS of a KI and direct investigations of the SS in YbB$_{12}$ have not been done extensively. An ARPES study on a clean (001) surface of this material revealed surface band dispersion consistent with what is expected from the SS of a TKI [61] but could not establish the existence of TSS with absolute certainty. The spin texture in the SS of YbB$_{12}$ also remains to be determined.

In this work, we report a PTS study on the (001) surface of the single crystalline KI YbB$_{12}$, in which we discuss our findings in comparison to the results from recent transport measurements [49] and the ARPES study [61]. We also compare YbB$_{12}$ against the better understood KI SmB$_6$ to see if the interaction between the bulk excitations and the surface states [33] exists similarly in the two. In our PTS study, YbB$_{12}$ exhibits linear conductance at low bias, a signature originating from the V-shaped Dirac fermion density of states (DOS) around the Dirac point. The overall shape is similar to SmB$_6$ except that the bias at which the linearity ends is much lower than the hybridization gap edges. Additionally, the conductance curves when Pb is superconducting, do not show any additional pronounced peak. Such a peak at 5 mV signified inelastic tunneling involving bulk excitons in SmB$_6$ [33]. Based on these signatures and other details, we discuss the topological nature of the surface states in the KI YbB$_{12}$.

## II. MATERIALS AND METHODS

YbB$_{12}$ single crystals were grown by the traveling solvent floating zone method [62]. The orientation of as-grown crystals was determined using the backscattered X-ray Laue diffraction technique and then they were cut along the (001) direction into rectangular parallelepiped shape. Three of such cut pieces were used for the experiments in this study. The dc resistance of the unpolished and polished crystals was measured using the standard four-probe method by attaching thin aluminum strips using silver paint.

A similar procedure for making tunnel junctions on single crystals of YbB$_{12}$ was followed as had been used for SmB$_6$ single crystals [33]. Each cut piece was embedded in a mold made of Stycast® epoxy (2850-FT) and then polished. Tunnel junctions were made repeatedly on the three crystals by repolishing them after each measurement was completed. Surface smoothness on the atomic scale, essential for high-quality tunnel junctions, is achieved by polishing with diamond lapping films in the range of 3 – 0.25 μm particle size. Atomic force microscopy of the polished surfaces reveals a peak-to-dip roughness less than 10 Å (see Supplemental Material [63]), showing that smooth and clean crystal surfaces are achieved after polishing. These polished crystals are etched with an argon ion beam in a high vacuum chamber to remove any contaminants on the surface and to obtain a boron-rich top layer, which is then plasma-oxidized in the same chamber to form boron oxide to serve as the tunnel barrier [64]. Since the characteristics of the tunnel barrier are extremely sensitive to both ion-beam etching and plasma oxidation processes, the etching/oxidation power and durations were optimized through multiple runs. The crystal edges are then painted with Duco® cement to ensure the electrical isolation on the edges and to define the junction areas, after which the counter-electrode is deposited by thermal evaporation. Typical junction dimensions are in the range of $(0.2 – 0.3) \times (0.1 – 0.7)$ mm². For the measurement of the differential conductance, thin aluminum strips are attached to both the bottom and top electrodes. The measurement is done by a standard four-probe lock-in technique as a function of temperature down to 1.75 K and magnetic field up to 14 T.

## III.   RESULTS AND DISCUSSION

The temperature dependence of the dc electrical resistance, R(T), is measured across the (001) surface of one of the $YbB_{12}$ crystals used in our PTS study and normalized against its resistance at room temperature, $R_N(T)$ as shown Fig. 2a. In order to see whether the crystal shows similar temperature dependence before and after the polishing process, the normalized dc resistance at both stages of the crystal is compared. The temperature dependence remains almost the same, attesting the robustness of the surface states against a harsh process like that of polishing, as is the case for $SmB_6$ [33]. The inset of Fig. 2a shows the R(T) for all the three crystals, indicating that it is similar in all the crystals including the increase by about five orders of magnitude, a typical insulating behavior. A closer inspection reveals that this increase in R(T) is not monotonic, suggesting that $YbB_{12}$ is not a trivial bulk insulator. More specifically, the slope of the R(T) curve increases below ~50 K, similar to $SmB_6$ [33]. This increase in the slope of the R(T) curve indicates more insulating character and therefore, this can be attributed to the hybridization gap opening in the bulk. This is followed by a broad hump in the range of 10 K – 30 K, reminiscent of a similar hump observed in $SmB_6$ from 15 K – 20 K [33]. A further increase in the slope is observed as the temperature is decreased below 15 K, showing a two-gap behavior. The magnitude of the two gaps is obtained (Fig. 2b) by the thermal activation model of resistivity ($\rho \propto e^{\frac{\Delta}{2kT}}$, where $\rho$ = resistivity, $\Delta$ = energy gap, k = Boltzmann constant, and T = temperature) and the gap values are found to be $\Delta_1$ = 10.95 meV (15 K < T < 40 K) and $\Delta_2$ = 5.11 meV (7 K < T < 15 K). While $\Delta_1$ is similar to those obtained in previous transport measurements, $\Delta_2$ is slightly larger ($\Delta_1$ = 10.9 – 11.2 meV and $\Delta_2$ = 3.33 – 4.23 meV, for the same temperature ranges [49]). The $\Delta_1$ value in YbB$_{12}$ is comparable to SmB$_6$ ($\Delta_1$ = 10.4 – 11.2 meV), whereas $\Delta_2$ is bit larger in SmB$_6$ ($\Delta_2$ = 5.8 – 6.6 meV) [65].

According to the theory [14], the metallic SS are expected to develop in close correspondence with the formation of the bulk gap in KIs. But experimentally, the opening of this bulk hybridization gap in SmB$_6$ is not immediately accompanied by the development of the conducting SS. This was speculated to be due to the presence of interactions between the SS and the bulk spin excitons, which become negligible only at a lower temperature where the spin excitons condense [33,65]. The first signatures of the conducting SS only appear when the crystals are further cooled down to about 3 – 4 K, where there is a rapid decrease in the slope of the R(T) curves [49] indicating the development of the conducting SS. While stoichiometric SmB$_6$ crystals exhibit a pronounced plateau below ~ 4 K [16-18,33], Sm-deficient crystals only show a decrease in the slope but no pronounced plateau down to the lowest measurement temperature [65]. In the present study, the YbB$_{12}$ crystals show a clear decrease in the slope at low temperature (~ 3 K), as seen in Fig. 2a (A plateau becomes more apparent upon further cooling [63]). Similar to SmB$_6$, this temperature is much lower than the bulk gap opening temperature and therefore this indicates the presence of interactions between the SS and the bulk spin excitons, as was speculated in the case of SmB$_6$ [33]. But transport measurements alone cannot confirm whether the plateau seen in our resistance data is due to the contribution from the TSS or in-gap states in the bulk arising from disorder. Spectroscopic measurements can help us better address these questions. We have employed PTS since it is a surface-sensitive technique with high energy resolution and momentum selectivity and, thus, well-suited for the study of the SS.

We have chosen Pb as the counter-electrode in our tunnel junctions because the sharpness of its superconducting features in the differential conductance, G(V) = dI/dV, serves as an important diagnostic for their quality. Fig. 3a shows the normalized conductance, G$_N$(V), obtained by dividing G(V) by the value at the negative maximum bias, for two junctions on the same crystal. The sharp Pb superconducting features, i.e., the pronounced coherence peaks and the normalized zero-bias conductance (ZBC) being close to zero (~6.3 x $10^{-3}$), confirm their high quality. Both conductance curves are similar in their overall shape. However, they are different in some detailed features including the sharpness of coherence peaks and the depths of the ZBC, which can be due to a slight difference in the microstructure of the junction area. Fig. 3b compares the sharpness of the Pb superconducting features observed in different junctions prepared with different parameters for the ion-beam etching and plasma oxidation. Similar sensitiveness of the junction quality to such processing parameters was also reported for SmB$_6$ [64], wherein it was speculated that the layer of disrupted boron octahedra resulting from the ion beam etching process might be too thin if the ion beam energy is low or it becomes inhomogeneous if the ion beam energy is high. The asymmetry of the coherence peaks is highly reproducible across junctions irrespective of the details of the barrier formation, which originates from the intrinsic DOS of the YbB$_{12}$, as seen below. Amongst the two junctions in Fig. 3a which show the lowest ZBC, the junction labeled as J1 is of higher quality as evidenced by more prominent coherence peaks. Plotted in Fig. 3c is the conductance taken after Pb is driven normal by an applied magnetic field. Further increasing the field well above the Pb critical field shows no change at low bias, but a slight increase at higher bias, whose origin is not currently clear [63].

Another standard diagnostic for the quality of the tunnel barrier is fitting the G(V) data in the high-bias region, which may contain only the barrier effect, to the Brinkman-Dynes-Rowell (BDR) model [66]. According to this model, such background conductance exhibits quadratic dependence on the bias voltage as follows:

$$G(V) = 3.16 \times 10^{10} H^{1/2} d e^{-1.025 d H^{1/2}} \times [1 - 0.171 \, d\Gamma H^{-3/2} V + 0.525 d^2 H^{-1} V^2], \qquad (1)$$

where $d$ is the barrier thickness (in Å), $H$ is the barrier height (in eV), and $\Gamma$ is the asymmetry of the barrier (in eV). As shown in Fig. 4, the fitting for J1 is good giving reasonable values for the fitting parameters: $d$ =18.49 Å, $H$ = 3.25 eV, and $\Gamma$ = 1.19 eV. This indicates that the tunnel barrier formed with optimized processing parameters for the ion beam etching and plasma oxidation is of a desirable shape. On the other hand, changing the processing parameters slightly result in a poor tunnel barrier (S2) with very different fitting parameters ($d$ = 24.85 Å, $H$ = 2.02 eV, and $\Gamma$ = 3.13 eV) in which the barrier height is less than the barrier asymmetry. This further attests the high sensitivity of the barrier properties to the processing parameters [64].

Normalizing the G(V) when the Pb is in a superconducting state against that obtained when the Pb is driven to a normal state at the same temperature by applying a magnetic field of 0.1 T removes the features intrinsic to $YbB_{12}$. This helps to study additional attributes in the conductance curves arising from other channels, e.g., involving inelastic tunneling [33,65]. Such a normalized G(V) curve is shown in Fig. 5a. It does not exhibit any additional features like the peak near +5 mV observed in $SmB_6$ [33]. This suggests that in this compound the interaction of the bulk spin excitons with the SS is not obvious down to the lowest measurement temperature (1.75 K), unlike in the case of $SmB_6$. On the other hand, while the asymmetry in the relative height of the coherence peaks in Fig. 5a may seem to be pointing towards the presence of such an interaction between the SS and the bulk excitons as in $SmB_6$ [33], the slight difference from $SmB_6$ in their temperature evolution raises questions on the origin of this asymmetry. This can be seen from the temperature evolution of the coherence peaks in Fig. 5b, where the asymmetry between the two peaks remains almost constant till they disappear near the critical temperature ($T_c$ = 7.25 K) of Pb. This is unlike $SmB_6$, where the asymmetry in the relative height of the coherence peaks increases with increasing temperature until they vanish at the $T_c$ of Pb [33]. These features, which are absent in $YbB_{12}$ data, were attributed to the inelastic tunneling involving bulk spin excitons in $SmB_6$ [33], suggesting that the SS interact with the bulk excitons. Their absence in the present conductance spectra indicates that the topological nature of the SS in $YbB_{12}$ is very different from that of $SmB_6$ as similar studies on $SmB_6$ indicate that such interactions between the surface state and the bulk excitons alter the topological protection of the surface state drastically [33,40]. Measuring the conductance at even lower temperatures and the second harmonics may provide further information with regards to the presence of bulk excitations in $YbB_{12}$ and their interactions with its SS.

The V-shaped conductance at low bias, as expected for the Dirac fermion DOS, is clearly seen in Fig. 6a when the Pb is driven to a normal state. This linearity starts tapering off slowly outside the ± 2 mV range. This is a very narrow region compared to the size of the bulk gap found by other studies (~15 mV) [53]. As already discussed regarding the temperature dependence of the electrical resistance, this non-correlation between the emergence of the SS and the bulk hybridization gap, also seen in SmB$_6$ [33], can be explained by the interaction of the SS with the bulk spin excitons resulting in the incoherence of the SS. In spite of this similarity in the low-bias feature with SmB$_6$ [33], two major differences can be seen on a closer inspection. Firstly, the kink-hump structure seen in the SmB$_6$ data [33] is absent in YbB$_{12}$. While the kink-hump structure in SmB$_6$ was associated with elastic tunneling involving bulk spin excitons with which the SS interact, the absence of this structure in the YbB$_{12}$ spectra does not rule out such an interaction either, as already suggested by the non-correlation between the emergence of these SS and the bulk hybridization gap. This indicates that the possible interaction of the SS with bulk spin excitons is manifested very differently in YbB$_{12}$, possibly due to the difference in the exact topological nature of the YbB$_{12}$ surface from SmB$_6$. Secondly, the linear conductance at low bias shows only one slope. This indicates that the YbB$_{12}$ (001) surface has only one kind of Dirac band, unlike in the SmB$_6$ (001) surface [33]. This is supported by theoretical band calculations [46], which indicate that four Dirac points appear at the four M points on the (001) surface of YbB$_{12}$ and hence, four Dirac bands of the same type contribute to the low bias conductance shown in Fig. 6a. While the qualitative similarity in the low-bias conductance between YbB$_{12}$ and SmB$_6$ is an evidence for the topological origin of the SS in YbB$_{12}$, in agreement with other experiments [49,61], the minute differences between the two discussed in this section indicate that the SS in these two materials are not exactly the same. Band calculations also indicate that YbB$_{12}$ is a much more strongly correlated insulator than SmB$_6$ and that YbB$_{12}$ is a topological crystalline insulator, unlike SmB$_6$ which is a $Z_2$ topological insulator [46].

The temperature dependence of G(V) in the low-bias region (with the Pb driven normal) is shown in Fig. 6b. The linearity starts to vanish as the temperature is increased above ~3 K, which agrees with the corresponding change in the resistance slope shown in Fig. 2a. This further strongly ties the origin of the resistance slope change to the TSS. Also, the G(V) does not show any additional feature, e.g., a peak arising from a trivial origin like an impurity band. The temperature evolution in Fig. 6b is used to obtain the normalized ZBC at different temperatures, which is plotted in Fig. 6c. A turning point can be seen around 4 K, below which these topological surface states can be said to develop and therefore dominantly contribute to the conductance [33,65]. In contrast to SmB$_6$ [33], the normalized ZBC in YbB$_{12}$ does not exhibit clear plateauing down to the lowest measurement temperature (1.75 K), in line with the resistance behavior (Fig. 2).

Figures 7a and 7b show the temperature evolution of G(V) over a much wider bias range for two junctions prepared on different crystals whose surfaces were processed under slightly different conditions. Figure 7c shows the corresponding temperature evolution of the ZBC in these two junctions. While Fig. 7a is for a junction with sharp Pb superconducting features, Fig. 7b is for a junction with significantly smeared Pb features. At low temperature, both junctions exhibit a characteristic hump structure around -50mV. Interestingly, it consists of far more pronounced double humps in Fig. 7b, which is unexpected since this junction shows more smeared Pb coherence peaks than the other junction as mentioned above. As the temperature is increased, this double-hump structure gets thermally smeared, resulting into a single hump, and finally vanishing around 220 K. The curves in Fig. 7a also seem to become parabolic above ~200 K. The prominence of the double-hump structure appears uncorrelated with the sharpness of Pb coherence peaks, as seen in all measured junctions [63], but always appear in the negative bias region and approximately in the same bias voltage range. The double-hump structure across different junctions is compared in Fig. 7d. Their temperature dependence is also similar across all these junctions indicating a common origin. A similar double-hump structure in the negative bias region was also seen in $SmB_6$ [34] around -25 mV. The band calculations in $YbB_{12}$ show two bands very close to each other near the X point [46] around the same energy range where the double-hump structure is observed in our data. These bands coincide at the $\Gamma$ point and start separating as the X point is approached. Since in our PTS study the major tunneling direction is [001], that is, along the $\Gamma$-X direction (Fig. 1b), we speculate that the double-hump structure may originate from these two closely spaced bands near the X point. We further speculate that the variation in its prominence may have to do with the detailed surface microstructure as the prominence is comparable among the junctions in each sample. Further investigation is required for better explanation of this observation.

IV. **SUMMARY**

In summary, we have obtained tunneling conductance spectra of the (001) surface of $YbB_{12}$ single crystals and analyzed the data in comparison to $SmB_6$. Similar to $SmB_6$, the linear conductance obtained at low bias strongly suggests the existence of surface Dirac fermions. The linearity consists of a single slope pointing towards the existence of just one kind of Dirac band on the (001) surface in agreement with results from previous theoretical calculations. The temperature dependence of the low bias conductance reveals that the contribution from the SS becomes evident only at very low temperatures (< 4 K) in agreement with the low-temperature resistance behavior. Another similarity to $SmB_6$ is the double-hump structure seen in the negative bias region, which has been explained again by theoretical band calculations.

We observe that the temperature at which the metallic SS start dominantly contributing to the conductance is not coincident with the much higher bulk gap opening temperature (~ 50 K). Furthermore, the bias up to which the linearity of the conductance exists (± 2 mV) is very low compared to the bulk gap (~15 mV) suggesting that the SS might lose the coherence well before reaching the band edges. We speculate that, just like in $SmB_6$, this discrepancy from what's expected theoretically may be caused by a possible interaction between the SS and the bulk spin excitons. However, such an interaction is manifested in $YbB_{12}$ very differently as evidenced by the absence of any clear elastic or inelastic tunneling signatures involving these bulk excitons down to the lowest measurement temperature (1.75 K). This difference from $SmB_6$ qualitatively suggests that even though the SS in $YbB_{12}$ is of topological origin, its exact nature is different from that of $SmB_6$. Obtaining conductance spectra at lower temperatures corresponding to the clear plateau in the resistivity will help us better understand the interaction between the SS and the bulk excitons. Further theoretical analysis is also required to arrive at a more comprehensive picture of the topological nature of the SS in $YbB_{12}$.

**ACKNOWLEDGEMENTS**

The work at NHMFL and FSU was supported by the NSF/DMR-2003405, NSF/DMR-1644779, and the State of Florida. The work at JHU was funded by the DOE/BES EFRC DE-SC0019331, JHU Catalyst Fund, and NSF/DMR(PARADIM)-2039380.

# REFERENCES

[1] M. Z. Hasan and C. L. Kane, *Colloquium: Topological insulators*, Rev. Mod. Phys. **82**, 3045 (2010).

[2] X. L. Qi and S. C. Zhang, *Topological insulators and superconductors*, Rev. Mod. Phys. **83**, 1057 (2011).

[3] M. Z. Hasan and J. E. Moore, *Three-Dimensional Topological Insulators*, Annu. Rev. Condens. Matter Phys. **2**, 55 (2011).

[4] L. Fu, C. L. Kane, and E. J. Mele, *Topological insulators in three dimensions*, Phys. Rev. Lett. **98**, 106803 (2007).

[5] B. A. Bernevig, T. L. Hughes, and S. C. Zhang, *Quantum spin Hall effect and topological phase transition in HgTe quantum wells*, Science **314**, 1757 (2006).

[6] M. Konig, S. Wiedmann, C. Brune, A. Roth, H. Buhmann, L. W. Molenkamp, X. L. Qi, and S. C. Zhang, *Quantum spin hall insulator state in HgTe quantum wells*, Science **318**, 766 (2007).

[7] X. Dai, T. L. Hughes, X. L. Qi, Z. Fang, and S. C. Zhang, *Helical edge and surface states in HgTe quantum wells and bulk insulators*, Phys. Rev. B **77**, 125319 (2008).

[8] H. J. Zhang, C. X. Liu, X. L. Qi, X. Dai, Z. Fang, and S. C. Zhang, *Topological insulators in Bi2Se3, Bi2Te3 and Sb2Te3 with a single Dirac cone on the surface*, Nat. Phys. **5**, 438 (2009).

[9] Y. Xia *et al.*, *Observation of a large-gap topological-insulator class with a single Dirac cone on the surface*, Nat. Phys. **5**, 398 (2009).

[10] Y. L. Chen *et al.*, *Experimental Realization of a Three-Dimensional Topological Insulator, Bi2Te3*, Science **325**, 178 (2009).

[11] M. Dzero, J. Xia, V. Galitski, and P. Coleman, *Topological Kondo Insulators*, Annu. Rev. Condens. Matter Phys. **7**, 249 (2016).

[12] H. Tsunetsugu, M. Sigrist, and K. Ueda, *The ground-state phase diagram of the one-dimensional Kondo lattice model*, Rev. Mod. Phys. **69**, 809 (1997).

[13] P. S. Riseborough, *Heavy fermion semiconductors*, Adv. Phys. **49**, 257 (2000).

[14] M. Dzero, K. Sun, V. Galitski, and P. Coleman, *Topological Kondo Insulators*, Phys. Rev. Lett. **104**, 106408 (2010).

[15] J. W. Allen, *Foreward for special issue of philosophical magazine on: topological correlated insulators and SmB6*, Philos. Mag. **96**, 3227 (2016).

[16] D. J. Kim, J. Xia, and Z. Fisk, *Topological surface state in the Kondo insulator samarium hexaboride*, Nat. Mater. **13**, 466 (2014).

[17] P. Syers, D. Kim, M. S. Fuhrer, and J. Paglione, *Tuning Bulk and Surface Conduction in the Proposed Topological Kondo Insulator SmB6*, Phys. Rev. Lett. **114**, 096601 (2015).

[18] Y. S. Eo *et al.*, *Comprehensive surface magnetotransport study of SmB6*, Phys. Rev. B **101**, 155109 (2020).

[19] G. Li *et al.*, *Two-dimensional Fermi surfaces in Kondo insulator SmB6*, Science **346**, 1208 (2014).

[20] B. S. Tan *et al.*, *Unconventional Fermi surface in an insulating state*, Science **349**, 287 (2015).

[21] Z. Xiang, B. Lawson, T. Asaba, C. Tinsman, L. Chen, C. Shang, X. H. Chen, and L. Li, *Bulk Rotational Symmetry Breaking in Kondo Insulator SmB6*, Phys. Rev. X **7**, 031054 (2017).

[22] S. M. Thomas, X. X. Ding, F. Ronning, V. Zapf, J. D. Thompson, Z. Fisk, J. Xia, and P. F. S. Rosa, *Quantum Oscillations in Flux-Grown SmB6 with Embedded Aluminum*, Phys. Rev. Lett. **122**, 16640 (2019).

[23] J. D. Denlinger, J. W. Allen, J.-S. Kang, K. Sun, J.-W. Kim, J. H. Shim, B. I. Min, D.-J. Kim, and Z. Fisk, *Temperature dependence of linked gap and surface state evolution in the mixed valent topological insulator SmB6*, arXiv:1312.6637.

[24] E. Frantzeskakis *et al.*, *Kondo Hybridization and the Origin of Metallic States at the (001) Surface of SmB6*, Phys. Rev. X **3**, 041024 (2013).

[25] J. Jiang *et al.*, *Observation of possible topological in-gap surface states in the Kondo insulator SmB6 by photoemission*, Nat. Commun. **4**, 3010 (2013).

[26] C. H. Min, P. Lutz, S. Fiedler, B. Y. Kang, B. K. Cho, H. D. Kim, H. Bentmann, and F. Reinert, *Importance of Charge Fluctuations for the Topological Phase in SmB6*, Phys. Rev. Lett. **112**, 226402 (2014).

[27] H. Miyazaki, T. Hajiri, T. Ito, S. Kunii, and S. Kimura, *Momentum-dependent hybridization gap and dispersive in-gap state of the Kondo semiconductor SmB6*, Phys. Rev. B **86**, 075105 (2012).

[28] M. Neupane *et al.*, *Surface electronic structure of the topological Kondo-insulator candidate correlated electron system SmB6*, Nat. Commun. **4**, 2991 (2013).

[29] N. Xu *et al.*, *Direct observation of the spin texture in SmB6 as evidence of the topological Kondo insulator*, Nat. Commun. **5**, 4566 (2014).

[30] M. M. Yee, Y. He, A. Soumyanarayanan, D.-J. Kim, Z. Fisk, and J. E. Hoffman, *Imaging the Kondo insulating gap on SmB6*, arXiv:1308.1085.

[31] S. Rossler, T. H. Jang, D. J. Kim, L. H. Tjeng, Z. Fisk, F. Steglich, and S. Wirth, *Hybridization gap and Fano resonance in SmB6*, Proc. Natl. Acad. Sci. U.S.A. **111**, 4798 (2014).

[32] W. Ruan, C. Ye, M. H. Guo, F. Chen, X. H. Chen, G. M. Zhang, and Y. Y. Wang, *Emergence of a Coherent In-Gap State in the Sm B-6 Kondo Insulator Revealed by Scanning Tunneling Spectroscopy*, Phys. Rev. Lett. **112**, 136401 (2014).

[33] W. K. Park, L. N. Sun, A. Noddings, D. J. Kim, Z. Fisk, and L. H. Greene, *Topological surface states interacting with bulk excitations in the Kondo insulator SmB6 revealed via planar tunneling spectroscopy*, Proc. Natl. Acad. Sci. U.S.A. **113**, 6599 (2016).

[34] L. Sun, D. J. Kim, Z. Fisk, and W. K. Park, *Planar tunneling spectroscopy of the topological Kondo insulator SmB6*, Phys. Rev. B **95**, 195129 (2017).

[35] X. H. Zhang, N. P. Butch, P. Syers, S. Ziemak, R. L. Greene, and J. Paglione, *Hybridization, Inter-Ion Correlation, and Surface States in the Kondo Insulator SmB6*, Phys. Rev. X **3**, 011011 (2013).

[36] F. Lu, J. Z. Zhao, H. M. Weng, Z. Fang, and X. Dai, *Correlated Topological Insulators with Mixed Valence*, Phys. Rev. Lett. **110**, 096401 (2013).

[37] V. Alexandrov, M. Dzero, and P. Coleman, *Cubic Topological Kondo Insulators*, Phys. Rev. Lett. **111**, 226403 (2013).

[38] T. Takimoto, *SmB6: A Promising Candidate for a Topological Insulator*, J. Phys. Soc. Jpn. **80**, 123710 (2011).

[39] W. T. Fuhrman *et al.*, *Interaction Driven Subgap Spin Exciton in the Kondo Insulator SmB6*, Phys. Rev. Lett. **114**, 036401 (2015).
[40] G. A. Kapilevich, P. S. Riseborough, A. X. Gray, M. Gulacsi, T. Durakiewicz, and J. L. Smith, *Incomplete protection of the surface Weyl cones of the Kondo insulator SmB6: Spin exciton scattering*, Phys. Rev. B **92**, 085133 (2015).
[41] A. Arab, A. X. Gray, S. Nemsak, D. V. Evtushinsky, C. M. Schneider, D. J. Kim, Z. Fisk, P. F. S. Rosa, T. Durakiewicz, and P. S. Riseborough, *Effects of spin excitons on the surface states of SmB6: A photoemission study*, Phys. Rev. B **94**, 235125 (2016).
[42] J. Kim, C. Jang, X. F. Wang, J. Paglione, S. Hong, J. Lee, H. Choi, and D. Kim, *Electrical detection of the surface spin polarization of the candidate topological Kondo insulator SmB6*, Phys. Rev. B **99**, 245148 (2019).
[43] L. J. Kong *et al.*, *Spin-polarized surface state transport in a topological Kondo insulator SmB6 nanowire*, Phys. Rev. B **95**, 235410 (2017).
[44] P. Hlawenka *et al.*, *Samarium hexaboride is a trivial surface conductor*, Nat. Commun. **9**, 517 (2018).
[45] G. Akopov, M. T. Yeung, and R. B. Kaner, *Rediscovering the Crystal Chemistry of Borides*, Adv. Mater. **29**, 1604506 (2017).
[46] H. M. Weng, J. Z. Zhao, Z. J. Wang, Z. Fang, and X. Dai, *Topological Crystalline Kondo Insulator in Mixed Valence Ytterbium Borides*, Phys. Rev. Lett. **112**, 016403 (2014).
[47] L. Fu, *Topological Crystalline Insulators*, Phys. Rev. Lett. **106**, 106802 (2011).
[48] T. H. Hsieh, H. Lin, J. W. Liu, W. H. Duan, A. Bansil, and L. Fu, *Topological crystalline insulators in the SnTe material class*, Nat. Commun. **3**, 982 (2012).
[49] Y. Sato *et al.*, *Topological surface conduction in Kondo insulator YbB12*, J. Phys. D Appl. Phys. **54**, 404002 (2021).
[50] Z. Xiang *et al.*, *Quantum oscillations of electrical resistivity in an insulator*, Science **362**, 65 (2018).
[51] Z. J. Xiang *et al.*, *Unusual high-field metal in a Kondo insulator*, Nat. Phys. **17**, 788 (2021).
[52] H. Liu, M. Hartstein, G. J. Wallace, A. J. Davies, M. C. Hatnean, M. D. Johannes, N. Shitsevalova, G. Balakrishnan, and S. E. Sebastian, *Fermi surfaces in Kondo insulators*, J. Phys.-Condens. Mat. **30**, 16LT01 (2018).
[53] H. Okamura, T. Michizawa, T. Nanba, S. Kimura, F. Iga, and T. Takabatake, *Indirect and direct energy gaps in kondo semiconductor YbB12*, J. Phys. Soc. Jpn. **74**, 1954 (2005).
[54] T. Susaki *et al.*, *Low-energy electronic structure of the Kondo insulator YbB12*, Phys. Rev. Lett. **77**, 4269 (1996).
[55] M. Okawa, Y. Ishida, M. Takahashi, T. Shimada, F. Iga, T. Takabatake, T. Saitoh, and S. Shin, *Hybridization gap formation in the Kondo insulator YbB12 observed using time-resolved photoemission spectroscopy*, Phys. Rev. B **92**, 161108(R) (2015).
[56] J. M. Mignot, P. A. Alekseev, K. S. Nemkovski, L. P. Regnault, F. Iga, and T. Takabatake, *Evidence for short-range antiferromagnetic fluctuations in Kondo-insulating YbB12*, Phys. Rev. Lett. **94**, 247204 (2005).
[57] J. M. Mignot, P. A. Alekseev, K. S. Nemkovski, E. V. Nefeodova, A. Rybina, L. P. Regnault, N. Y. Shitsevalova, F. Iga, and T. Takabatake, *Neutron scattering study of spin and lattice dynamics in $YbB_{12}$*, Physica B **383**, 16 (2006).

[58] F. Iga, K. Suga, K. Takeda, S. Michimura, K. Murakami, T. Takabatake, and K. Kindo, *Anisotropic magnetoresistance and collapse of the energy gap in Yb1-xLuxB12*, J. Phys. Conf. Ser. **200**, 012064 (2010).

[59] T. T. Terashima, A. Ikeda, Y. H. Matsuda, A. Kondo, K. Kindo, and F. Iga, *Magnetization Process of the Kondo Insulator YbB12 in Ultrahigh Magnetic Fields*, J. Phys. Soc. Jpn. **86**, 054710 (2017).

[60] T. T. Terashima, Y. H. Matsuda, Y. Kohama, A. Ikeda, A. Kondo, K. Kindo, and F. Iga, *Magnetic-Field-Induced Kondo Metal Realized in YbB12*, Phys. Rev. Lett. **120**, 257206 (2018).

[61] K. Hagiwara *et al.*, *Surface Kondo effect and non-trivial metallic state of the Kondo insulator YbB12*, Nat. Commun. **7**, 12690 (2016).

[62] W. A. Phelan *et al.*, *On the Chemistry and Physical Properties of Flux and Floating Zone Grown SmB6 Single Crystals*, Sci. Rep. **6**, 20860 (2016).

[63] Online Supplemental Material (hyperlink to be inserted by the publisher, http://link.aps.org/supplemental/xxxx).

[64] J. A. Sittler and W. K. Park, *Self-oxidation-formed boron oxide as a tunnel barrier in SmB6 junctions*, J. Alloy Compd. **874**, 159841 (2021).

[65] W. K. Park, J. A. Sittler, L. H. Greene, W. T. Fuhrman, J. R. Chamorro, S. M. Koohpayeh, W. A. Phelan, and T. M. McQueen, *Topological nature of the Kondo insulator SmB6 and its sensitiveness to Sm vacancy*, Phys. Rev. B **103**, 155125 (2021).

[66] W. F. Brinkman, R. C. Dynes, and J. M. Rowell, *Tunneling Conductance of Asymmetrical Barriers*, J. Appl. Phys. **41**, 1915 (1970).

[67] Y. Sato, *Quantum Oscillations and Charge-Neutral Fermions in Topological Kondo Insulator YbB$_{12}$* (Springer Singapore, 2021).

**FIGURE CAPTIONS**

**Figure 1.**

a. **Crystal structure of KI YbB$_{12}$ (adapted from [67]).** It has a face-centered cubic (FCC) structure similar to NaCl. The Yb ions (blue) make up the FCC lattice like Na and the B$_{12}$ cubooctahedra (green) takes up the Cl positions.

b. **Bulk and (001) surface Brillouin zones of YbB$_{12}$ (adapted from [46]).** The (001) surface has been studied by PTS in this study.

**Figure 2.**

a. **DC resistance normalized against room temperature resistance before and after polishing the crystal. Inset – resistance of all the three YbB$_{12}$ crystals.** The change in slope at around 50 K and the plateauing below 3 K are marked by arrows.

b. **Arrhenius plot of the resistance in 2(a).** The linear fit of the low temperature data shows a two-gap behavior: $\Delta_1$=10.95 meV (15 K < T < 40 K) and $\Delta_2$=5.11 meV (7 K < T < 15 K).

**Figure 3.**

a. **Normalized G(V) for the best junctions with Pb as counter-electrode.**
b. **Comparison of G(V) of junctions on different crystal surfaces processed using different conditions.** The quality of the junction is characterized by the sharpness of the coherence peaks and the depth of the zero-bias dip.
c. **Magnetic field dependence of normalized G(V).** There is no change in G(V) as the field is increased above 0.1 T. Below 0.1 T, the superconducting features of Pb are visible.

**Figure 4.**

a. **BDR fit for the high bias region of G(V) for two typical junctions used for our study.** The J1 barrier parameters are found to be d=18.49 Å, H=3.25 eV, $\Gamma$=1.19 eV. These values show that the tunnel barrier has a desirable shape. On the other hand, slightly different processing parameters result in S2, whose barrier parameters are found to be d=24.85 Å, H=2.02 eV, and $\Gamma$=3.13 eV. These values indicate poorer barrier quality.

**Figure 5.**

a. **G(V) with Pb in superconducting state normalized against G(V) with Pb in normal state.** No obvious features can be observed arising due to the interaction between Pb DOS and YbB$_{12}$ surface electrons.

b. **Temperature evolution of the coherence peaks.** The asymmetry between the coherence peaks stays constant as the temperature is changed. The peaks vanish once the Pb is in a normal state.

**Figure 6.**

a. **Normalized G(V) at low bias showing V-shaped DOS as expected for Dirac DOS.** The linearity is similar on both sides of zero bias and tapers of below -2.08 mV and above 2.18 mV. The red dashed lines have been drawn as aid to the eye.

b. **Temperature evolution of normalized G(V) at low bias.** The curves have been shifted for shifted for clear visibility. The linearity gradually becomes less prominent as the temperature is increased and vanishes above 3K.

c. **Normalized ZBC vs T derived from Fig. 6b.** There is a turning point seen at around 4 K due to the surface state contribution. At the lowest measurement temperature of 1.75 K, there is a clear sign of plateauing signifying that the surface state contribution dominates below this temperature.

**Figure 7.**

a. **Temperature dependence of G(V) for junction S2.** The double hump structure around 50 mV is smeared and appears as a single hump. The curves become parabolic above ~200 K.

b. **Temperature dependence of G(V) for junction S4.** The double hump structure around 50 mV can be clearly seen. As the temperature is increased, the double hump turns into a single hump and then vanishes above ~220 K.

c. **ZBC vs T for the junctions S2 and S4.**

d. **Normalized G(V) curves for various junctions zoomed around the double hump structure.**

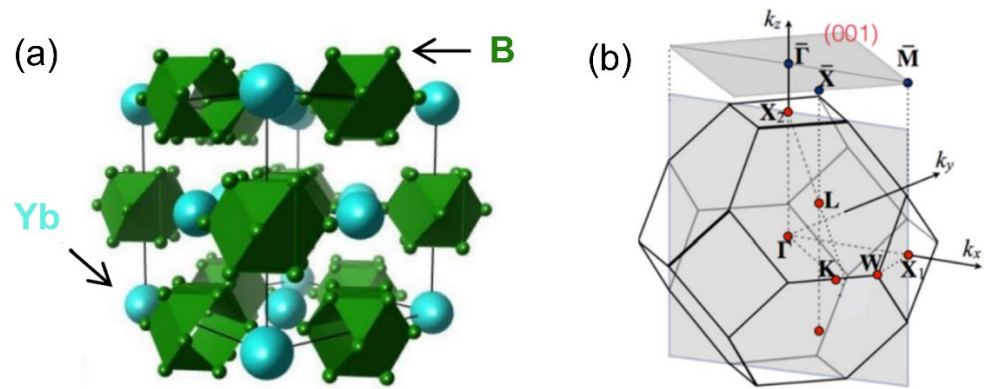

**Fig. 1, Gupta et al.**

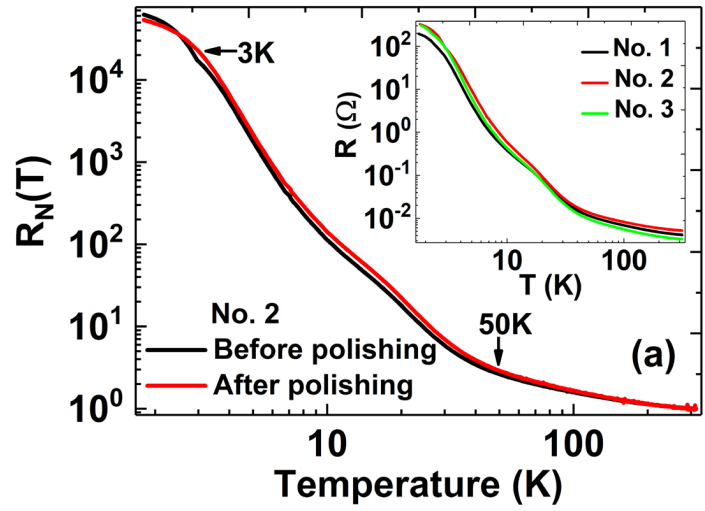

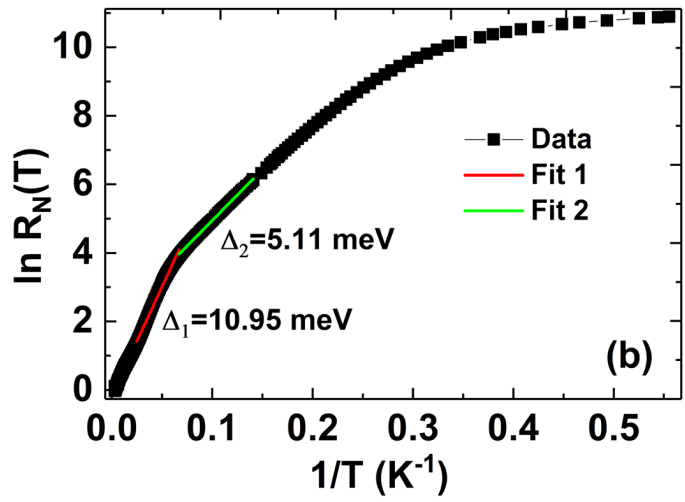

**Fig. 2, Gupta et al.**

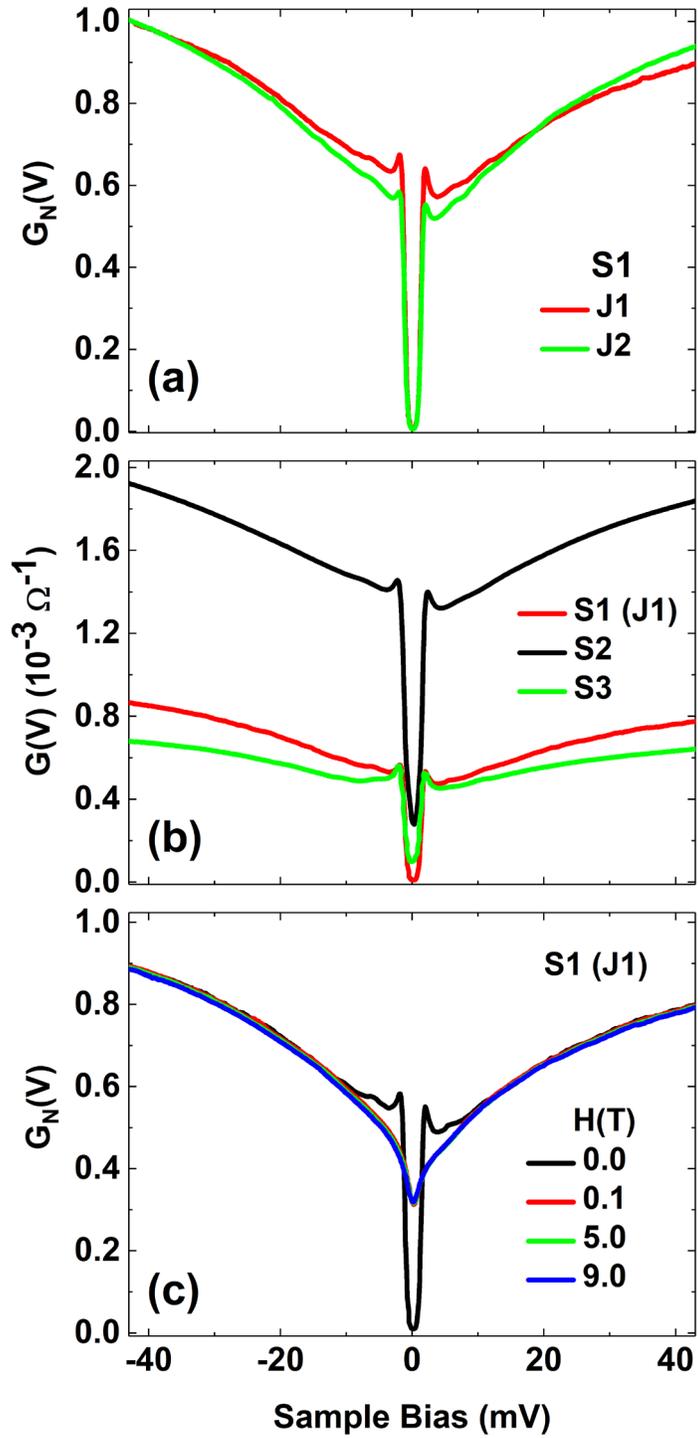

Fig. 3, Gupta et al.

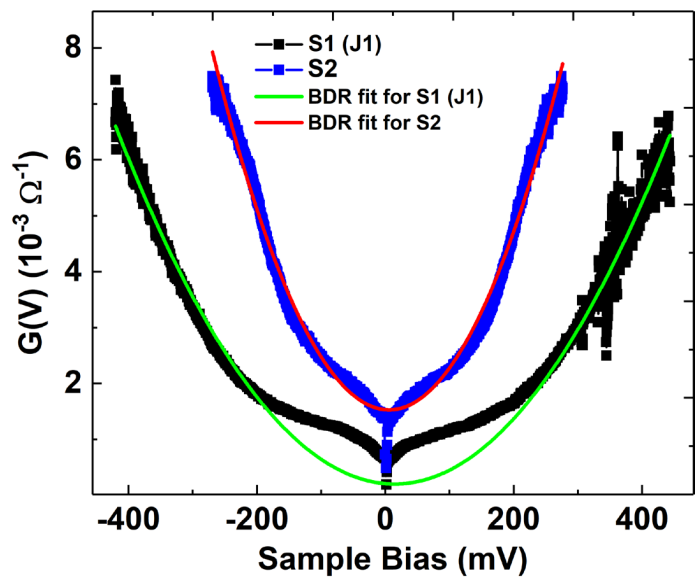

**Fig. 4, Gupta et al.**

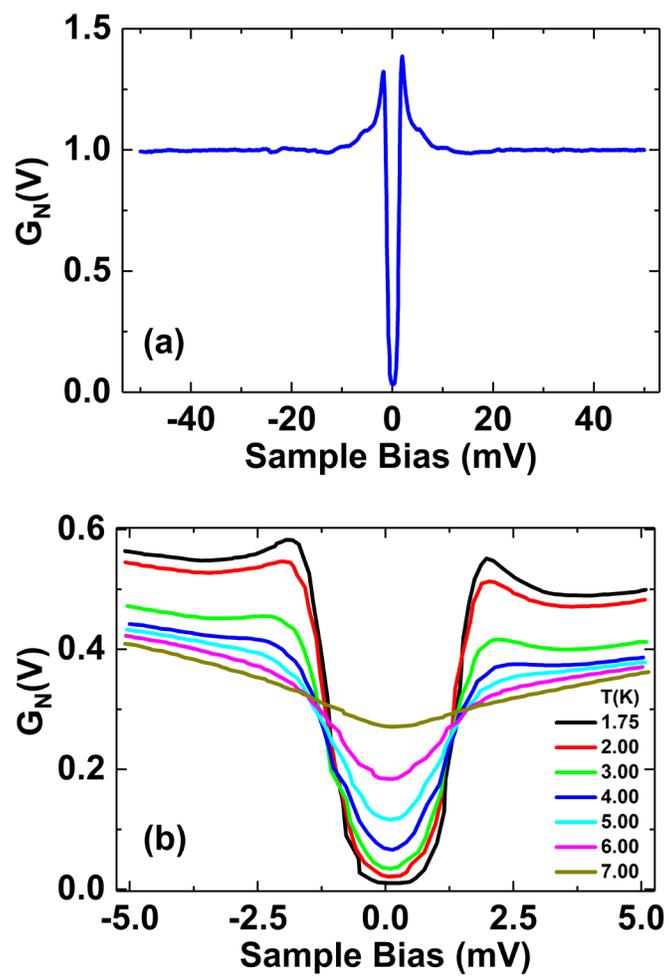

**Fig. 5, Gupta et al.**

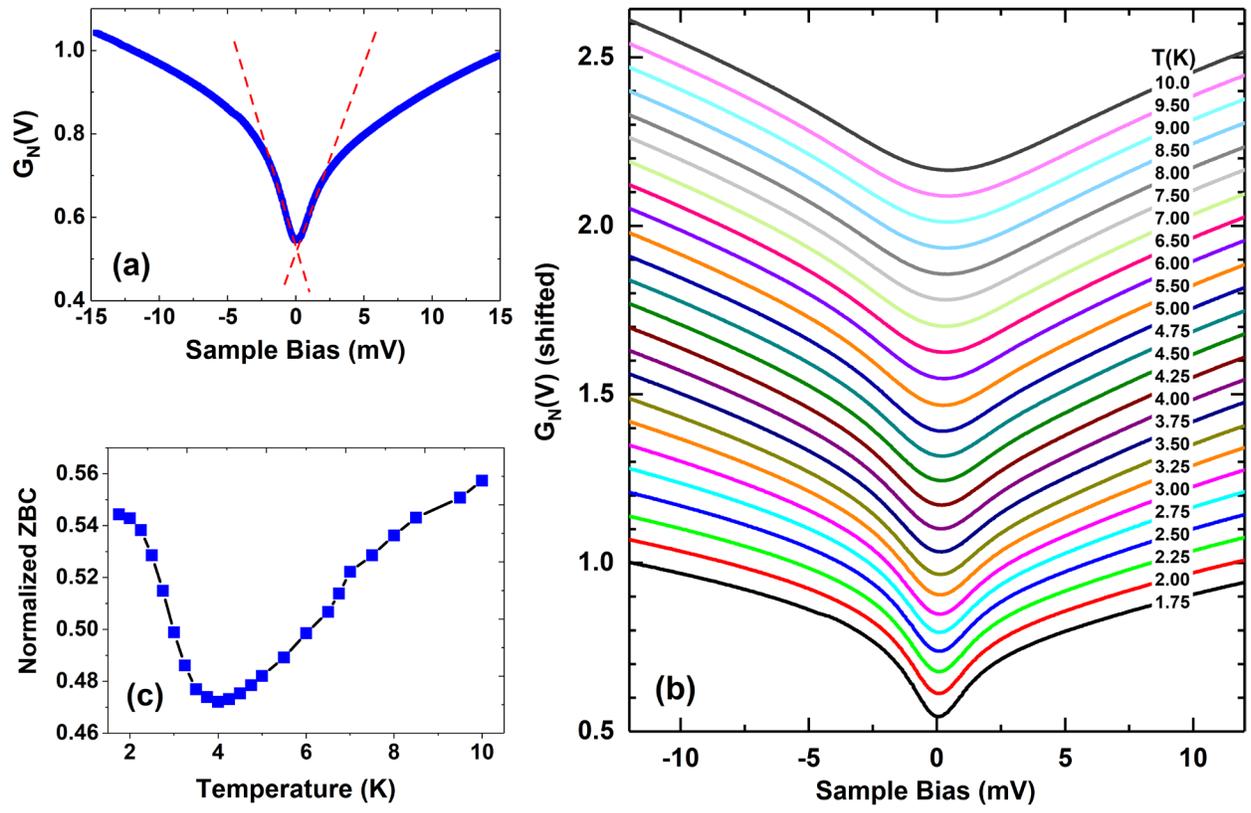

Fig. 6, Gupta et al.

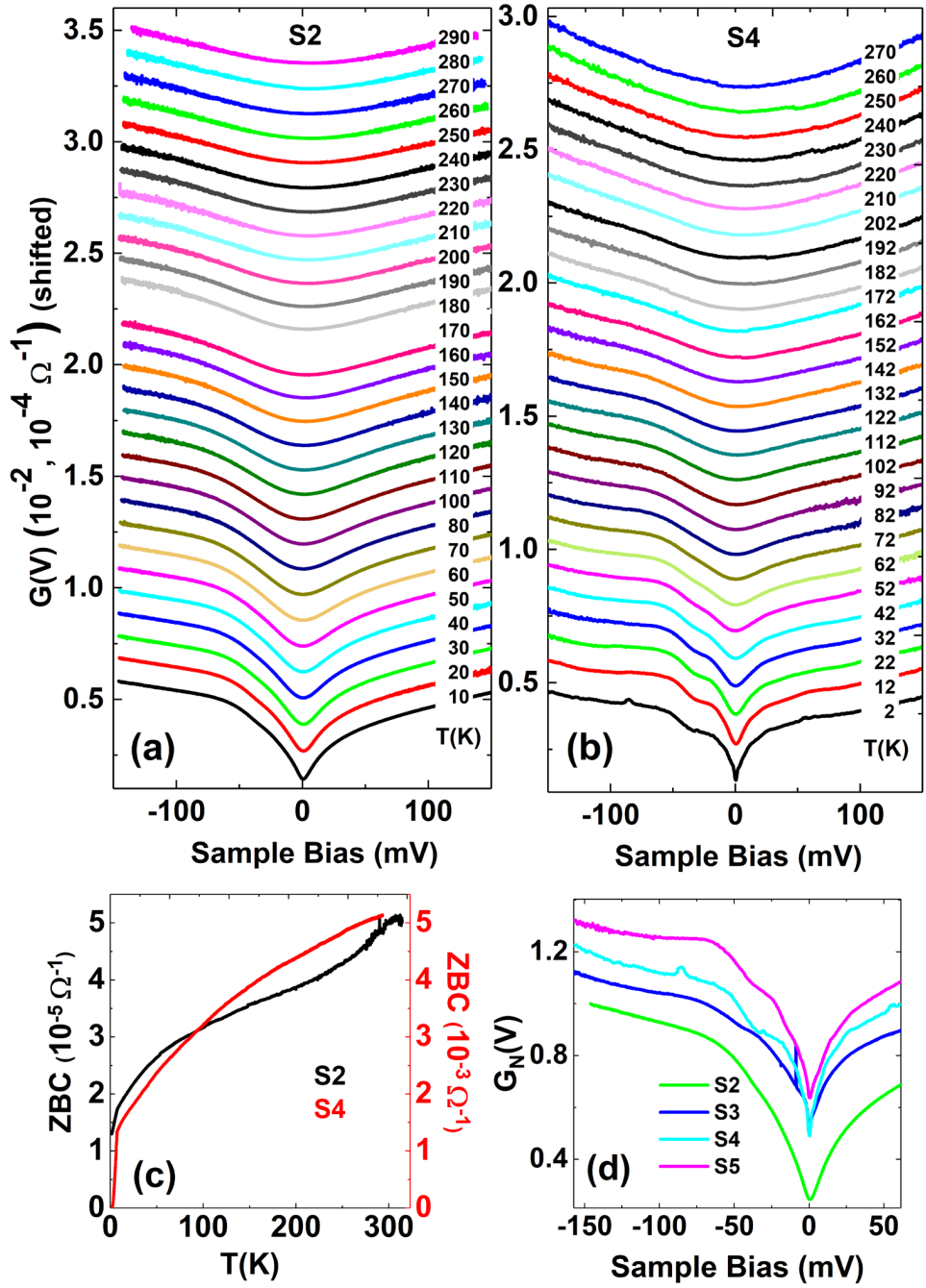

**Fig. 7, Gupta et al.**

# Supplemental Material for

# Topological surface states in the Kondo insulator YbB$_{12}$ revealed via planar tunneling spectroscopy

A. Gupta[1,2], A. Weiser[3], L. H. Greene[1,2], L. Pressley[4,†], Y. Luo[4], C. Lygouras[4], J. Trowbridge[5], W.A. Phelan[4,5,6,‡], C.L. Broholm[4], T. McQueen[4,5,7], W. K. Park[1,*]

[1]National High Magnetic Field Laboratory, Florida State University, Tallahassee, FL 32310, USA

[2]Department of Physics, Florida State University, Tallahassee, FL 32306, USA

[3]Department of Physics, Astronomy, Geology, and Environmental Science, Youngstown State University, Youngstown, OH 44555, USA

[4]Department of Physics and Astronomy, Johns Hopkins University, Baltimore, MD 21218, USA

[5]Department of Chemistry, Johns Hopkins University, Baltimore, MD 21218, USA

[6]Hopkins Extreme Materials Institute, Johns Hopkins University, Baltimore, MD 21218, USA

[7]Department of Materials Science and Engineering, Johns Hopkins University, Baltimore, MD 21218, USA

*Corresponding author: wkpark@magnet.fsu.edu

†Present address: Oak Ridge National Laboratory, Oak Ridge, TN 37830, USA

‡Present address: Los Alamos National Laboratory, Los Alamos, Mail Stop E574, Los Alamos, NM 87545, USA

# 1. Characterization of the polished YbB$_{12}$ crystal surfaces

As mentioned in the main text, clean crystal surfaces with atomic-scale smoothness are essential to fabricate high-quality tunnel junctions. We analyzed polished surfaces of the YbB$_{12}$ single crystals used in our PTS study using an atomic force microscope (AFM). They are found to be extremely smooth with average roughness of less than 5 Å over 10 × 10 µm$^2$ area, as shown in Fig. S1a, and peak-to-dip roughness less than 10 Å along a 10 µm-long cross-sectional line, as shown in Fig. S1b.

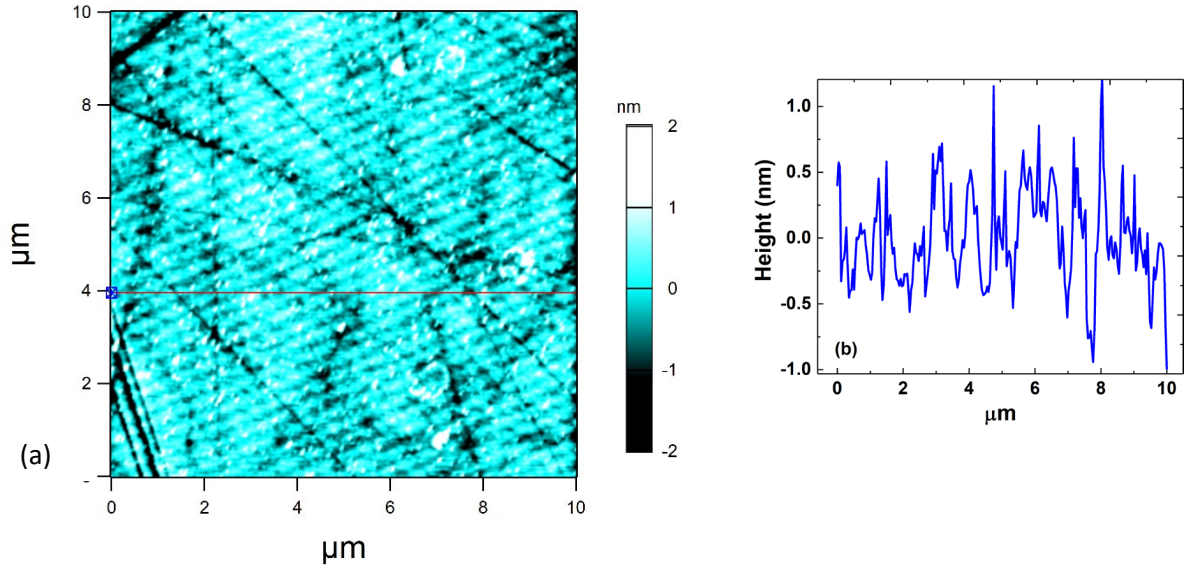

**Figure S1. (a) AFM image of a typical YbB$_{12}$ crystal surface after the whole polishing and cleaning procedure.** The red line indicates the line over which the cross-sectional profile is shown in Fig. S1b. The periodic pattern along the diagonal direction is an experimental artefact. The average rms roughness of this surface is 3.7 Å. (b) **Cross-sectional line profile along the red line marked in Fig. S1a.** The peak-to-dip roughness along this line is less than 10 Å.

# 2. Comparison of R(T) curves between YbB$_{12}$ and SmB$_6$

Figure S2 compares the temperature evolution of dc resistance normalized against room temperature resistance, R$_N$(T), of a YbB$_{12}$ single crystal and an SmB$_6$ single crystal used in a previous study [1]. The YbB$_{12}$ data was taken at John Hopkins University down to a dilution fridge temperature, showing a clear plateau below ~3 K. Both curves are qualitatively similar to each other, but they vary in their details. Near room temperature, both curves coincide and as the temperature is decreased, YbB$_{12}$ starts increasing faster than that of SmB$_6$. Both show a change in slope around 50 K, following which there is a broad hump. While in SmB$_6$ the hump appears in the range of 15 K - 20 K, the hump in YbB$_{12}$ is broader (10 K – 30 K). The latter can be more clearly seen in Fig. 2a in the main text. The major difference lies in the development of the low-temperature plateau. While the plateauing begins around the same temperature (~3 K) in both compounds, the plateau in SmB$_6$ develops much more quickly than in YbB$_{12}$, in which the development is much smoother in comparison. Thus, while we could investigate the tunneling conductance in the plateau region at the lowest measurement temperature (1.75 K) for SmB$_6$, we could not do the same for YbB$_{12}$ and it is necessary to use a Helium-3 fridge to investigate the same in the future.

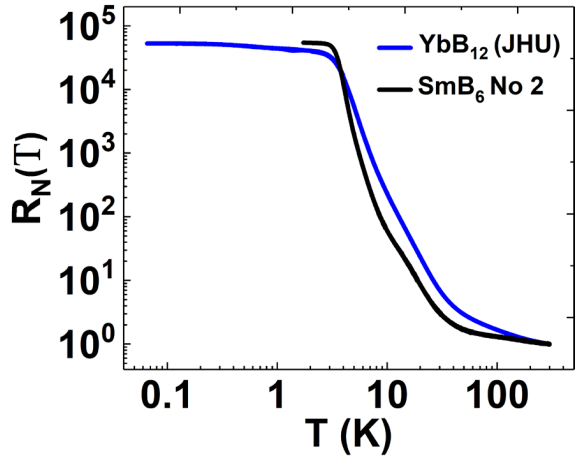

**Figure S2. DC electrical resistance normalized against room temperature resistance for YbB$_{12}$ and SmB$_6$ single crystals.** The YbB$_{12}$ resistance is measured down to a lower temperature to show clear resistance plateau which is seen in SmB$_6$ at a higher temperature.

### 3. Magnetic Field Dependence of G(V)

Figure S3 shows the tunneling conductance G(V) measured under various applied magnetic fields up to 9 T. While we see in Fig. 3c in the main text that there is no change in the G(V) at higher fields as compared to the zero-field data in junction J1, this stands true only at lower bias. At higher bias, the conductance continuously increases as the applied field is increased. This behavior is rather unexpected since G(V) at higher bias largely reflects the barrier properties. Certainly, there

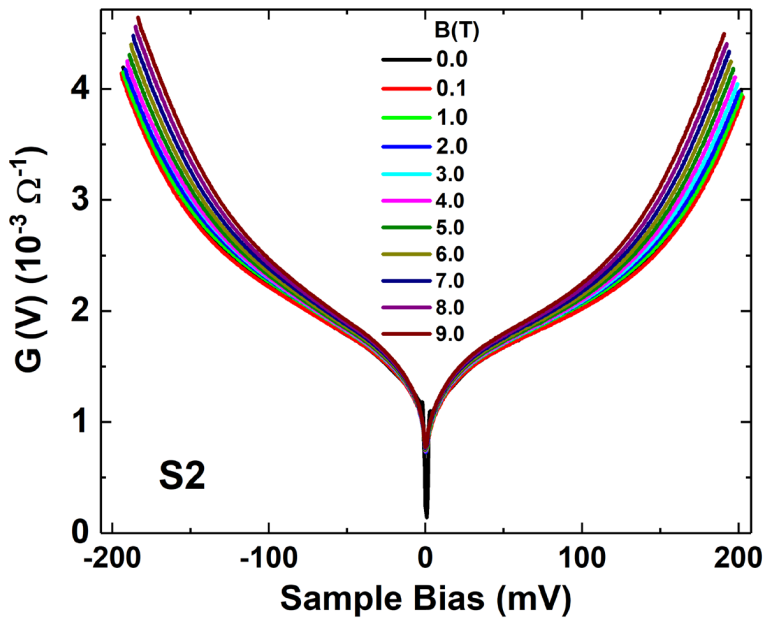

**Figure S3. Tunneling conductance G(V) as a function of bias voltage for various applied magnetic fields for junction S2.** The field was applied perpendicularly to the junction. While the curves coincide at very low bias, G(V) increases with

is no reason why the tunnel barrier would change at high fields. Therefore, further investigations are needed to understand the magnetic field dependence of the tunneling conductance in YbB$_{12}$.

## 4. Relation between the prominence of the double hump in G(V) and Pb superconducting signatures

As described in the main text, the barrier properties are quite sensitive to the ion-beam etching and plasma oxidation parameters and the optimum processing parameters result in the best quality junction evidenced by sharp Pb superconducting features. Interestingly, the double-hump structure in the negative bias region is reproducibly observed around a similar bias range irrespective of the processing parameters used. This behavior nominally suggests that the prominence of the double-hump seems to be independent of the junction quality. This is clearly seen in Fig. S3, where the non-correlation between the sharpness of Pb superconducting features and the prominence of the double hump structure is shown by comparing the sharpness of the Pb superconducting features for the four junctions shown in Fig. 7d in the main text.

By comparing these junctions, we see that S2 has sharp Pb superconducting features but smeared double-hump structure. On the other hand, S3 has even sharper Pb features but it has a clear double-hump structure. Both S4 and S5 show pronounced double-hump structures. While the Pb superconducting signatures in S4 are largely suppressed, they are better in S5 but still diminished in comparison to S3 and S2. This attests that there is no correlation between the quality of the junction and the prominence of the double-hump structure. We speculate that the detailed microstructure in the junction area may play a role here.

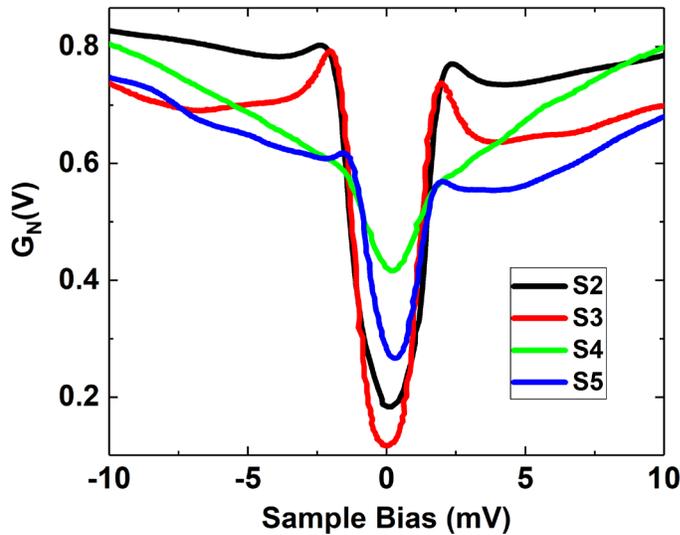

**Figure S4. Normalized G(V) for the four junctions whose double hump structures were compared in Fig. 7d in the main text.** The Pb superconducting signatures in S2 and S3 are good, largely suppressed in S4, and somewhat smeared in S5.

## REFERENCES

[1] W. K. Park, L. N. Sun, A. Noddings, D. J. Kim, Z. Fisk, and L. H. Greene, P Natl Acad Sci USA **113**, 6599 (2016).